\documentclass[11pt]{article}
\usepackage{epsfig}
\usepackage{times}
\usepackage{a4wide}
\newcommand{\sss}{\vspace{.2in}}

\newcommand{\be}{\begin{equation}}
\newcommand{\ee}{\end{equation}}
\newcommand{\bea}{\begin{eqnarray}}
\newcommand{\eea}{\end{eqnarray}}

\begin{document}
\begin{flushright}
~\hfill{\footnotesize IP/BBSR/01-12}\\[-4pt]
~\hfill{\footnotesize ISN-01-48}\\[-4pt]
~\hfill{\footnotesize {\tt quant-ph/0105130}}\\[-4pt]
~\hfill{\footnotesize \today}
\end{flushright}
\sss
\sss
\begin{center}
{\Large {\Large \bf Testing Hall-Post Inequalities \\[6pt]
With Exactly Solvable $\mathbf{N}$-Body Problems}}
\end{center}
\vspace{.5in}

\def\vec#1{{\bf #1}}
\begin{center}
{Avinash {\sc Khare}$^{(1)}$ and Jean-Marc {\sc Richard}$^{(2)}$}

\vspace{.3in}
$^{(1)}$ Institute of Physics,\\[-2pt]
Sachivalaya Marg, Bhubaneswar 751005,India\\

$^{(2)}$Institut des Sciences Nucl\'eaires,
Universit\'e Joseph Fourier--CNRS-IN2P3\\[-2pt]
53, avenue des Martyrs,F 38026 Grenoble, France.
\end{center}
\sss
\sss
\begin{abstract}
The Hall--Post inequalities provide lower bounds on $N$-body energies
in terms of $N'$-body energies with $N'<N$. They are rewritten and
generalized to be tested with exactly-solvable models of
Calogero-Sutherland type in one and higher dimensions. The bound for
$N$ spinless fermions in one dimension is better saturated at large
coupling than for noninteracting fermions in an oscillator
potential.
\end{abstract}
\newpage
It is important to obtain good upper and lower bounds on the binding
energy of $N$-particle systems. Since upper bounds are provided by
variational estimates, it is natural that a major emphasis in recent
years has been on obtaining good lower bounds.

The Hall--Post inequalities consist of lower bounds to $N$-body
energies in terms of $N'$-body energies with $N'<N$ and modified
constituent masses or coupling constants.
\cite{hp,bmr,fr}. Applications of these inequalities have been
proposed for studying the thermodynamic limit of large systems and the
stability of matter \cite{fr}, or the relation of baryons to mesons in
hadron spectroscopy \cite{bmr,art}.

  So far, the Hall-Post inequalities have been tested in great detail
for few-body systems, for which accurate numerical calculations can be
performed, or in the large $N$ limit, within some approximation. It
seems appealing to test the inequalities for arbitrary $N$ with
energies which are calculated exactly. The purpose of this note is to
adapt and apply the Hall--Post inequalities to several
exactly-solvable $N$-body models.

Inspired by the
seminal work of Calogero and Sutherland, in recent years, ground state
energy and at least a part of the excitation spectrum (if not the full
spectrum) has been obtained for several new $N$-body problems. Some of
these Hamiltonians  have already found applications in a variety of
areas. 

We broadly consider four types of $N$-body problems.

\begin{enumerate}

\item $N$-body problems in one dimension with two-body (and also may 
be
one-body) interaction in the presence of either two-body or one-body 
oscillator
potential. Some examples of this class are $A_{N-1}$ \cite{ca}, 
$BC_N$, $D_N$
\cite{op} models. The case of $D$ dimensions is considered in 
\cite{pkg}.

\item $N$-body problems in one dimension but with periodic boundary 
conditions
and with two-body (and also possible one-body interaction). Again, 
some
examples of this class are the $A_{N-1}$ \cite{su} and $BC_N$, $D_N$
\cite{op} models.

\item $N$-body problems in one dimension with two-body inverse square
interaction in the presence of a hyper-Coulomb
potential \cite{ak}.

\item $N$-body problems of Calogero-type in two and higher dimensions 
with
two and three-body interactions in the presence of one or two-body 
oscillator
potential \cite{cm,kr}.

\end{enumerate}

Our conclusions can be summarized as follows. While comparing the
first and second type of models with the Hall--Post inequalities, we
find that contrary to our initial belief, the bound for spinless
fermions in one dimension is better saturated in the large coupling
limit rather than for $N$-noninteracting fermions in an oscillator
potential.  For the hyper-Coulomb case, using convexity argument, we
derive a slightly better lower bound than was known before. Finally,
we show that Hall--Post bounds also work reasonably well in models 
with
both two and three-body interactions.

In this paper we shall consider models with $N$ identical particles of
mass $m$ (which we shall put equal to 1 without any loss of 
generality)
and whose interaction does not depend on their spins. This 
corresponds to
the Hamiltonian
\be\label{1a}
H_N (m, g_1, g_2, g_3) = \sum_{i=1}^{N} \frac{{\vec p}_i^2}{2m}
+g_1V(r_i) + g_2 \sum_{i<j} V(r_{ij}) + g_3 \sum_{i<j<k} V(r_{ijk})~.
\ee
For this case, it has been shown that the corresponding $N$-particle
bound-state energy $E_N$ satisfies the bound \cite{fr,art}
\be\label{1b}
E_N (m, g_1, g_2, g_3) \ge \frac{N}{N-1} E_{N-1} \left(m, 
g_1,\frac{N-1}{N-2}g_2,
\frac{N-2}{N-3}g_3\right)~.
\ee
In case the Hamiltonian (\ref{1a}) is translationally invariant, then 
this
bound can be improved yielding the new inequality \cite{hp,bmr}
\be\label{1c}
E_N (m, g_2, g_3) \ge \frac{N-1}{N-2} E_{N-1}\left 
(m,\frac{N}{N-1}g_2,
\frac{N(N-2)}{(N-1)(N-3)}g_3\right)~.
\ee

As a representative of the first type of models, we consider the 
original
Calogero problem \cite{ca} for which the $N$-body Hamiltonian is 
given by
($\hbar =m=1$)
\be\label{1}
H = -\frac{1}{2}\sum_{i=1}^{N} \frac{d^2}{dx_i^2}
+ \sum_{i<j=1}^{N} \bigg [\frac{\omega^2}{4} (x_i -x_j)^2
+ \frac{g}{(x_i-x_j)^2} \bigg ]~.
\ee
As shown by Calogero, the ground-state energy for this $N$-particle
system is given by
\be\label{2}
E_{N} = \sqrt{\frac{N}{8}}
\bigg [N^2 -1 + (\beta -1)N(N-1) \bigg ]\omega~,
\ee
where
\be\label{3} g = \beta (\beta -1), \quad\hbox{i.e.,}\quad \beta =
\frac{1}{2} \pm \frac{\sqrt{1+4g}}{2}~, 
\ee
while the corresponding
ground-state eigenfunction is given by 
\be\label{4} 
\psi =\prod_{i=1}^{N} (x_i -x_j)^{\beta}\,\times\,
\exp\bigg [-\frac{\omega}{\sqrt{2N}}\sum_{i<j=1}^{N} (x_i 
-x_j)^2\bigg ]~.  
\ee 
Note that we have to choose the positive square root in eq.~(\ref{3})
so that $g=0$ corresponds to $\beta =1$ and hence to fermions.

For translationally invariant Hamiltonians with two-body interaction,
as the one given in eq.~(\ref{1}), the bound (\ref{1c}) takes the form
\be
\label{5} E_N (m, \omega,g) \ge \frac{N-1}{N-2} E_{N-1} \left(m,
\omega\sqrt{\frac{N}{N-1}},\frac{N}{N-1}g\right)~.  
\ee

Using the bound-state energy expression (\ref{2}) as derived for the
Hamiltonian (\ref{1}), it is easily seen that 
\be\label{6} 
E_{N-1}\left(m, \omega\sqrt{\frac{N}{N-1}},\frac{N}{N-1}g\right) =
\sqrt{\frac{N}{8}} \bigg [N(N-2) + (\beta' -1)(N-1)(N-2) \bigg
]\omega~, 
\ee
where 
\be\label{7} 
\beta' = \frac{1}{2} +
\frac{\sqrt{N(2\beta-1)^2 -1}}{2\sqrt{N-1}}~.  
\ee 
Using eqs.~(\ref{2}), (\ref{5}) and (\ref{6}) we then get 
\be\label{8}
R_N\equiv\frac{E_N (m, \omega,g)}
{\frac{N-1}{N-2} E_{N-1}(m, 
\omega\sqrt{\frac{N}{N-1}},\frac{N}{N-1}g)} 
\ge
\frac{N+1+(\beta-1)N}{N+(\beta'-1)(N-1)}~.  
\ee
In the non-interacting fermion limit (i.e., $g =0$ or $\beta = \beta'
=1$), the right-hand side is $\frac{N+1}{N}$ while in the
strong-coupling limit (i.e., $g \rightarrow \infty$) the r.h.s.\ is
$\frac{\sqrt{N}}{\sqrt{N-1}}$.  On the other hand, as $N \rightarrow
\infty$ in one dimension, then the ratio goes to 1, for all values of
$g$. Thus we find that contrary to our naive expectation, the bound is
better satisfied by strongly interacting rather than non-interacting
fermions in an oscillator potential.

\begin{figure}
\begin{center}
\epsfxsize=0.5\linewidth
\epsfbox{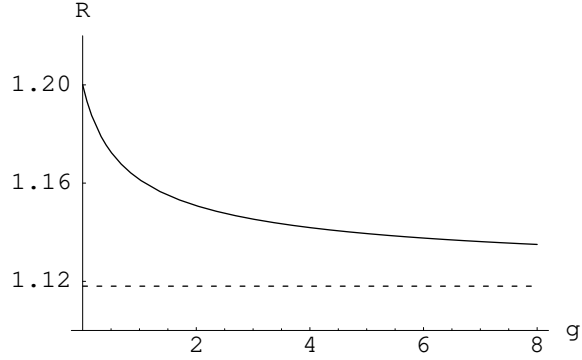}
\end{center}
\caption{\label{Fig1} Hall--Post ratio $R_N$, as defined in
eq.~(\ref{8}) for the Calogero model (\ref{1}), in the case of $N=5$
particles, as a function of the coupling constant $g$. The dotted line
is the $g\to\infty$ limit.}
\end{figure}

Fig.~\ref{Fig1} displays the ratio $R_N$ as a
function of the coupling constant $g$, in case $N=5$. One starts at 
$g=0$ from the value $R_5=6/5$ corresponding to a pure oscillator. 
Then the ratio evolves regularly towards its large $g$ limit 
$\sqrt{5/4}$.

We have also examined how Hall-Post inequalities  behave in 
similar models like $BC_N$, $D_N$, and even periodic models like
those of Sutherland. In all these cases we found a behaviour similar
to that given in Fig.~\ref{Fig1}. We might add here that many of these
models are not translationally invariant and instead of the lower
bound (\ref{5}), we have to use the simpler bound (\ref{1b}).

As a second example, we now consider a model of the third type
\cite{ak}, i.e.,
\be\label{9}
H = -\frac{1}{2}\sum_{i=1}^{N} \frac{d^2}{dx_i^2}
+ \sum_{i<j=1}^{N}
 \frac{g}{(x_i-x_j)^2}
-\frac{\alpha^2}{\sqrt{\sum_{i<j=1}^{N} (x_i -x_j)^2}}~,
\ee
which contains a hypercentral Coulomb interaction.
The $N$-particle ground-state binding energy has been shown to be
\be\label{10}
E_N = -\frac{\alpha^2}{N[N-2+N(N-1)\beta]^2}~,
\ee
with $\beta$ and $g$ being again related by eq.~(\ref{3}).

We now show that using a convexity argument, one can adapt the bound
(\ref{1c}) to this case. In particular, we use the fact that if a
function $f$ is a convex function, then
\be\label{11}
f\left(\frac{\alpha a + \beta b+\cdots }{\alpha + \beta+\cdots 
}\right)
\ge \frac{\alpha}{\alpha + \beta+\cdots } f(a)
+\frac{\beta}{\alpha + \beta+\cdots } f(b)+\cdots~,
\ee
for positive weight factors $\alpha$, $\beta$, \dots.
On using $f(x)=-1/\sqrt{x}$, it is easily shown that in this case
the Hall-Post inequality takes the form
\be\label{12}
E_N (m, g, \alpha) \ge \frac{N-1}{N-2} E_{N-1}\left(m, \frac{N}{N-1}g,
\frac{(N-1)\sqrt{(N-2)}}{N^{3/2}}\alpha\right)~.
\ee

\begin{figure}
\begin{center}\epsfxsize=0.5\linewidth
\epsfbox{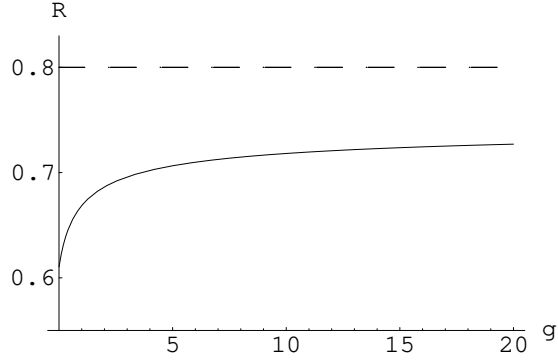}
\end{center}
\caption{\label{Fig2} Ratio $R_N$, as given in eq.~(13), for the model
of eq.~(\ref{9}), in the case of $N=5$ particles, as a function of the
coupling constant $g$. The dotted line is the $g\to\infty$ limit.}
\end{figure}

How good is this inequality? We can test that by using the energy 
eigenvalue
(\ref{10}) and computing $E_{N-1}$ for the appropriate couplings. On 
noting
the fact that the binding energy is negative, the Hall-Post inequality
(\ref{12}) in our case takes the form
\bea
\label{13}
&&R_N\equiv\frac{E_N(m, g, \alpha)}{\frac{N-1}{N-2} E_{N-1} (m, 
\frac{N}{N-1}g,
\frac{(N-1)\sqrt{(N-2)}}{N^{3/2}}\alpha)} \nonumber\\
&&\ \ \ \ \le \frac{(N)^2[N-3+(N-1)(N-2)\beta']^2}
{(N-1)^2[N-2+N(N-1)\beta]^2}~,
\eea
where
\be
\beta' = \frac{1}{2} + \frac{\sqrt{(N-1)(2\beta-1)^2 
-1}}{2\sqrt{N-2}}~.
\ee
We find that but for $N \rightarrow \infty$, $g$ arbitrary, this bound
is not as good as (\ref{8}) in the Calogero case.  In Fig.~\ref{Fig2},
we have plotted the ratio $R_N$ as a function of $g$ in case $N=5$
which gives an indication of how good this inequality is for finite 
$N$.

As the third and the last example, we consider a model of fourth type
\cite{cm,kr}  which is a Calogero-type model but in $D$-dimensions 
given by
\be\label{14}
H = \frac{1}{2}\sum_{i=1}^{N} {\vec p}_i^2
+ \frac{\omega^2}{4} \sum_{i<j=1}^{N} \vec{r}_{ij}^2
+ g \sum_{i<j=1}^{N} \frac{1}{{\vec r}_{ij}^2}
+G\sum_{{\scriptstyle i<j,k \atop \scriptstyle k\neq i,j}} 
\frac{{\vec r}_{ki}.{\vec r}_{kj}}{{\vec r}_{ki}^2 {\vec r}_{kj}^2}~,
\ee
where one also requires a three-body potential in addition to the
two-body potential. The $N$-boson ground-state energy in
%
\be\label{15} 
E_N\left(m,g,G(g)\right) =
\sqrt{\frac{N}{8}} \big [D(N-1)+N(N-1)\beta \big ]\omega~, 
\ee
provided the two-body coupling $g$ and the three-body coupling $G$ are
related by
\be\label{16} 
g = G +(D-2)\sqrt{G}~, \ G = \beta^2~,  
\ee 
this defining $G(g)$.  Note that unlike in one-dimension case, here
$\beta =0$ corresponds to $g = G= 0$ and hence to bosons
in a pure  oscillator potential.

Let us now examine how the Hall-Post inequality (\ref{1c}) fares in
this case.  It is not possible to test this inequality directly since
$g$ and $G$ have a specific relation $G(g)$ between them which in
general will not be satisfied if $g$ and $G$ are changed to
$\frac{N}{N-1}g$ and $\frac{N(N-2)}{(N-1)(N-3)}G$,
respectively. However, we make use of the fact that both the two and
three body terms in eq.~(\ref{14}) are positive in any dimension $D
(\ge 2$) and hence we can write an appropriate inequality. For example
we obviously have 
\bea
\label{17} 
E_N \left(m, \omega,g, G(g)\right) &\ge& \frac{N-1}{N-2}
E_{N-1}\left (m, \omega\sqrt{\frac{N}{N-1}},
             \frac{N}{N-1}g, \frac{N(N-2)}{(N-1)(N-3)}G(g)\right) 
\nonumber \\
&\ge& \frac{N-1}{N-2} E_{N-1}\left (m, \omega\sqrt{\frac{N}{N-1}},
                  \frac{N}{N-1}g, G(\frac{N}{N-1}g)\right)~, 
\eea
since
\be
\label{17a}
\frac{N(N-2)}{(N-1)(N-3)}G(g)\ge G(\frac{N}{N-1}g).
\ee

How good is this inequality? Using the exact $N$-particle binding 
energy
(\ref{15}) it is easily seen that
\be
\label{19}
\frac{E_N (m, \omega,g,G(g))}
{\frac{N-1}{N-2} E_{N-1} (m, 
\omega\sqrt{\frac{N}{N-1}},\frac{N}{N-1}g, G(\frac{N}{N-1}g))} 
 \ge \frac{D+N\beta}{D+(N-1)\beta'}~, 
\ee
where
\be\label{20}
\beta'=-\frac{D-2}{2}+\frac{\sqrt{N(2\beta+D-2)^2-(D-2)^2}}{2\sqrt{N-1}}~.
\ee
It is interesting that this bound is saturated for noninteracting
bosons ($\beta =0$ i.e., $g=G=0$) in an oscillator potential as well
as for large number of particles, no matter what the coupling is.  For
large coupling but finite $N$, the ratio is $\sqrt{N/(N-1)}$.  The
bound is also saturated at large $D$, as seen explicitly in
eq.~(\ref{19}).  This property of the large $D$ limit was noted by
Gonzalez-Garcia \cite{agg}.

Final remarks are in order:
\begin{enumerate}
\item
It would be interesting to extend the investigation to the case of
unequal masses. Some of the exactly-solvable models considered here
can be generalized as to accommodate different constituent masses
\cite{kof}.  The bound for unequal masses was discussed in the case of
$N=3$ or $N=4$ particles interacting through simple pairwise
potentials \cite{uneq}.
\item
Exactly solvable models are also available for particles with residual
interaction with nearest and next-to-nearest neighbour only
\cite{ajk}. It would be interesting to extend the Hall-Post
inequalities to this situation.
\item
One of the striking feature of some of the models considered here in
one or two dimensions is that the statistics evolves continuously as
one changes the coupling constant. Still, if one looks at the $E_N$ to
$E_{N-1}$ ratio measuring how far the inequalities are from
saturation, it evolves with a rather smooth and monotonic
behaviour. Fermions can approach saturation provided the coupling and
the number of particles are large enough.
\end{enumerate}

We hope to address some of the open issues in the near future.

\vskip .4cm\noindent
{\bf Acknowledgements}

AK would like to thank the members of the Institut des Sciences 
Nucl\'eaires,
Grenoble for warm hospitality during his trip there as a part of the
Indo-French Collaboration Project CEFIPRA 1501-1502, 
supported by the Indo-French Centre for Promotion of Advanced 
Research.
\newpage
%

\end{document}